\documentclass[12pt,preprint]{aastex}
\usepackage{color}
\definecolor{red}{rgb}{1,0.0,0.0}

\usepackage{amsmath} 
\usepackage{amssymb} 
\usepackage{graphics}
\usepackage{epsfig}  
\usepackage{float}
\def\be{\begin{equation}}
\def\ee{\end{equation}}
\def\ba{\begin{eqnarray}}
\def\ea{\end{eqnarray}}

\newcommand{\hMpc}{{\ifmmode{h^{-1}{\rm Mpc}}\else{$h^{-1}$Mpc }\fi}}  
\newcommand{\hGpc}{{\ifmmode{h^{-1}{\rm Gpc}}\else{$h^{-1}$Gpc }\fi}}  
\newcommand{\hmpc}{{\ifmmode{h^{-1}{\rm Mpc}}\else{$h^{-1}$Mpc }\fi}}  
\newcommand{\hkpc}{{\ifmmode{h^{-1}{\rm kpc}}\else{$h^{-1}$kpc }\fi}}  
\newcommand{\hMsun}{{\ifmmode{h^{-1}{\rm {M_{\odot}}}}\else{$h^{-1}{\rm{M_{\odot}}}$}\fi}}  
\newcommand{\hmsun}{{\ifmmode{h^{-1}{\rm {M_{\odot}}}}\else{$h^{-1}{\rm{M_{\odot}}}$}\fi}}  
\newcommand{\Msun}{{\ifmmode{{\rm {M_{\odot}}}}\else{${\rm{M_{\odot}}}$}\fi}}  
\newcommand{\msun}{{\ifmmode{{\rm {M_{\odot}}}}\else{${\rm{M_{\odot}}}$}\fi}}  
\newcommand{\bullb}{MACS J0025.4-1222}
\newcommand{\bulla}{1E0657---56} 
\shorttitle{Bullets in the MareNostrum}
\shortauthors{Forero-Romero et al.}

\begin{document} 

\title{Bullet Clusters in the MareNostrum Universe}
\author{Jaime E. Forero-Romero$^1$, Stefan Gottl\"ober$^1$ and Gustavo Yepes$^2$}
\affil{$^1$ Astrophysikalisches Institut Potsdam, An der Sternwarte 16, 14482
  Potsdam, Germany ; jforeo@aip.de\\
$^2$ Grupo de Astrof\'{\i}sica, Universidad Aut\'onoma de Madrid,   Madrid E-28049, Spain}

\begin{abstract}

We estimate the expected distribution of displacements between the dark
matter and gas cores in simulated clusters. We use the MareNostrum
Universe,  one of the  largest non radiative, SPH $\Lambda$CDM  cosmological
simulations.  We find that projected 2-D displacements between dark
matter and gas, equal or larger than the observed in the Bullet  Cluster, are
expected in $1\%$ to $2\%$ of the clusters with masses larger than $
10^{14}$ \hMsun. The 2-D displacement distribution is roughly the same
between redshifts $0<z<0.5$ when multiplied by a factor of $(1+z)^{-1/2}$.  We
conclude that the separations between dark matter and gas as observed
in the bullet cluster can be easily found in a $\Lambda$CDM
universe. Furthermore we find that the displacement distribution is
not very sensitive to  the normalization of the power
spectrum. Upcoming surveys could extend the measurements of these
displacements between dark matter and gas into large samples of
hundreds of clusters, providing a potential test for
$\Lambda$CDM.        

\end{abstract}


\section{Introduction}

The standard model of structure formation describes the Universe as a
three-component fluid (dark matter (DM), gas and radiation) plus a
cosmological constant.  There is a fundamental intrinsic difference
between the two matter components: while the DM can be modelled in a
very good approximation as collisionless fluid, the gaseous component
has an evolution dominated by collisional physics. As the
interaction between DM an baryons is only  through gravitational
forces, one can expect a decoupling between the components when the
collisional nature of baryons becomes important.

From the observational point of view, the decoupling is evident in
collisions of high mass clusters, observed, for the first time,  in the
cluster 1E0657-56 \citep{2006ApJ...648L.109C}, dubbed the Bullet
Cluster. For this object, a combined analysis of strong and weak
gravitational lensing and X ray observations gives a displacement of
around 200 kpc between the density peak of the gaseous and dark matter
components in the cluster.  More examples are given in
\citet{2007ApJ...668..781D, 2007ApJ...668..806M,
  2007ApJ...661..728J,2008ApJ...687..959B,2005ApJ...618...46J,2005ApJ...634..813J} 
but also see \citet{2008MNRAS.385.1431H} and the peculiar case in
\citet{2008ApJ...679L..81Q}.  The standard explanation for this
decoupling between DM and gas, is the collision between clusters of
galaxies. For example, in the case of a collision between two clusters
of comparable mass, the DM halos and galaxies go through each other
without falling apart from two body trajectories while the
intra-cluster medium, dominated by pressure effects, is stripped out of
the DM halos.    

In the Bullet Cluster, the expected configuration is that of a massive
sub-structure passing close to the center of the cluster, stripping away
the gas from the center of both structures, leaving two almost gas-naked cores
of dark matter and one single gas core in the middle. The velocity of
the expected collision could be inferred  observationally from the
temperature of the bow shock that is usually  present  in  the
collisions \citep{2003adu..book.....D}.  N-body simulations
(e.g.\citet{2007MNRAS.380..911S}, \citet{2008MNRAS.389..967M}) support
the model but  show also  that the  velocity estimation  should be
taken with care since the velocity of the encounter between dark matter
structures has large statistical and systematic uncertainties when it
is derived from the available observations.  

The Bullet Cluster has been recognized as potential stringent test for
models of structure formation.  On one hand it vindicated strongly the
need for a dark matter component, but on the other hand it raised
doubts on the likelihood of having such extreme events in a $\Lambda$CDM
universe \citep{2008MNRAS.384..343N}.  \citet{Hayashi06} attempted to
calculate the probability of finding large velocity collisions such as
the one guessed  for the Bullet Cluster. They used a large pure DM
N-body simulation (the Millennium Run, \cite{2005Natur.435..629S}) and
had to rely then on the information of the possible collisional
velocities of substructures inside a cluster sized DM halos. Taking
into account the difficulties in the interpretation of the data, the
observed velocities can be explained within the framework of
$\Lambda$CDM. Nevertheless, the frequency of such events changes from 1 to 5
in every 500 clusters,  depending on the assumptions  made on the
velocities of substructures. The lowest value corresponds to the
probability of having the sub-structure moving out of the center of the
cluster.    

The downside of \cite{Hayashi06} is that:

\begin{enumerate} 
\item The  mass assumed for the Bullet Cluster is lower than the most recent estimates. Using the
updated mass and the same velocity constraints, the Bullet probability
falls to $10^{-7}$ \citep{2007PhRvL..98q1302F}
 \item  Their  results
had to be extrapolated from the data in the Millennium simulation,
given that the abundance of clusters of comparable mass is not high
enough to perform  a more straightforward statistical study.
\end{enumerate}

 Both problems have been recently  addressed by  \cite{lee}. They use
 the MICE   dark matter only  simulation  with $\sim 200$ times larger
 volume than the Millennium Run. They use the same approach of finding
 large relative velocity of DM configurations. The large volume in the
 MICE  simulation allows them to gather a sample of clusters with
 masses similar to the  Bullet Cluster mass. Using the more stringent
 conditions on the collision velocity, they find that the probability
 of finding a system similar to the Bullet Cluster is between $3.3\times
 10^{-11}$ and $3.6\times 10^{-9}$.

Yet, there is another indisputable and clear example of a Bullet-like
configuration. The observations of the  cluster MACS J0025.4-1222
\citep{2008ApJ...687..959B} confirm the existence of two gas naked
matter cores and one single hot gas core. As opposed to the case of the
original Bullet Cluster, one cannot observe the X-ray emission from the
bow-shock. The derivation of a characteristic collision velocity, from
a simulation and available observations, can thus lead to confusing and
equivocal results.  In these two cases the only common and solid
observational physical characteristic    that can be measured is the
separation between the dominant gas clump  and the principal dark
matter clump.

It is expected  that not all the cases of Bullet like configurations,
will show a bow-shock X-ray emission, which can constrain the
sub-structure velocity \citep{shan}. Therefore, the results of the
analysis of dark matter only simulations cannot properly address the
likelihood of events similar to  the cluster \bullb.  If one wants to
use  the only robust and accessible observational feature, namely the
distance between the dominant DM and gas clumps, to test the $\Lambda$CDM
model predictions,  then one has to resort to hydrodynamical
simulations which include a description of the gas in  clusters.

Furthermore, as the number of observed clusters increases and will
increase even faster with upcoming surveys like eRosita, LSST and
Pan-STARRS, we are approaching an era in which a statistical study of
this decoupling would become possible through weak lensing and galaxy
number counts as a proxy for DM distribution and the X-ray emission as
a tracer of the gas component. As a first step in this direction
\citet{shan}, studied a compilation of 38 clusters.  They found 13
objects in which the separation between the center of mass of the gas
clump and of the dark matter halo is greater than 50 kpc and 3 more
clusters in which the separation is greater than 200 kpc.

The purpose of this paper is to aboard the likelihood of this
decoupling from the point of view of the distance between the DM and
gas cores in clusters measured in one of the largest volume
hydrodynamical simulations done up to now: the so-called {\em
  MareNostrum Universe} \citep{2007ApJ...664..117G}.

The structure of the paper is as follows: In Section
\ref{sec:simulation} we present the main features of the simulations,
in Section \ref{sec:algorithm} we define the algorithm to find the
separation between gas and DM structures. In Section
\ref{sec:proba} we present the probability of finding a cluster with a
given projected displacement on the sky. We inspect in Section
\ref{sec:power} the dependence of that probability with respect to the
normalization of the initial power spectrum.  In Section
  \ref{sec:challenge} we scrutinize whether the Bullet Cluster should
  be considered as a challenge to $\Lambda$ CDM in the context of our
  study. We summarize our results and state our conclusions in
Section \ref{sec:conclusions}. 

\section{Simulation}
\label{sec:simulation}

The analysis presented in this paper uses mainly the Marenostrum
Universe SPH cosmological simulation described in
\citet{2007ApJ...664..117G}. This non-radiative simulation
 was run using the  GADGET2 TreePM+SPH code
\citet{2005MNRAS.364.1105S} and follows the  evolution of
gas and dark matter from z=40 to z=0 in a comoving cube of $500$ \hMpc
on a side. The cosmology used  corresponds to  the spatially flat
concordance model with the
following parameters:  the total mass density $\Omega_m=0.3$, the baryon
density $\Omega_b=0.045$, the cosmological constant $\Omega_{\Lambda}=0.7$, the Hubble
parameter $h=0.7$, the slope of the initial power spectrum $n=1$ and
the normalization $\sigma_8=0.9$.  The number of particles used for each of
the DM and gas component was $1024^3$, resulting in a mass of $8.3 \times
10^9 h^{-1} M_{\odot}$ for the DM particles and $1.5 \times 10^9 h^{-1} M_{\odot}$
for the gas particles.  The spatial force resolution was set to an
equivalent Plummer gravitational softening of 15 \hkpc comoving.  

In order to study resolution effects on the distance
determination, we have also used a lower resolution simulation run from the
same initial conditions as the MareNostrum Universe. Here, the number of
particles is $2\times 512^3$ and the gravitational smoothing length is
correspondingly $\epsilon \sim 25 $ \hkpc.  We have performed two
more simulations with the same number of particles but different
cosmological parameters (equivalent to those obtained from fitting the
3rd year of WMAP data \citep{2007ApJS..170..377S})
  and  two different power
spectrum normalizations, a low $\sigma_8 = 0.75$ and a medium
$\sigma_8 = 0.8$.  The objective is to
quantify the effect of that cosmological parameter on the separation
probabilities.  In Table \ref{table:sims} we summarize the
main features of  our simulations. More details of these simulations can be found elsewhere
\citep{2007ApJ...666L..61Y}.

We analyse in the following a non-radiative simulation of a large
cosmological volume. It is well known that this approximation fails if
one considers the inner regions of clusters (within about 0.1 of the
virial radius). However, the outer regions of clusters are nearly
self-similar and different simulations agree in a remarkably successful
description of the temperature profile in the outside regions of
clusters. There are many studies of the formation of isolated clusters
which take into account cooling processes and star formation, however
these simulations still fail to describe the inner structure of
clusters correctly. Most probably additional processes, like AGN
feedbacks,  should be taken into account for a proper description. In
a recent review,  Borgani and Kravtsov \citep{2009arXiv0906.4370B}
discuss in great detail the issue of cosmological simulations of
galaxy clusters. In any case, the statistical study of cluster mergers
needs a large simulation volume  and large number of particles to
properly resolve the merger physics. This task  is not yet  possible
to be done  when  radiative physics is also considered. 
Furthermore, there are studies that suggest that the incomplete and
rough non-radiative  description provides a fair approximation to the
dynamics of the gas peaks  \citep{2006MNRAS.373..881P}.

\begin{table}
\begin{center} 
\begin{tabular}{cccccc}\hline\hline
Name & Particles & $\sigma_{8}$ &  $\Omega_M$& $\Omega_\Lambda $ & $n$ \\\hline
MN-1024  & $2 \times 1024^3$ & 0.9 & 0.30  & 0.70  & 1\\
MN-512  & $2 \times 512^3$ &  0.9 & 0.30  & 0.70 & 1 \\
MN-512-MS  & $2 \times 512^3$ &  0.8 &0.24  &0.73 & 0.95 \\
MN-512-LS  & $2 \times 512^3$ & 0.75 & 0.24  & 0.73 & 0.95 \\\hline
\end{tabular}
\caption{Summary of the main characteristics of the four simulations
  used in this work.}
\label{table:sims}
\end{center}
\end{table}

\begin{table}
\begin{center}
\begin{tabular}{ccc}\hline\hline
Redshift & \# Selected Clusters & Most massive cluster \\
z &  & ($10^{14}$ \hMsun)\\\hline
0 & 4063 & 25\\
0.3 & 2662 & 20\\
0.5 & 1826 & 13\\\hline
\end{tabular}
\caption{Analyzed redshifts $z$, number of selected clusters with
  $M > 10^{14}$ \hMsun and  mass of the most massive cluster at
  the given redshift in the simulation volume}
\label{table:lista}
\end{center}
\end{table}

\begin{table*}
\begin{center}
\begin{tabular}{cccccc}\hline\hline
Mass bin& Minimum Mass & Maximum Mass & \# Halos  & \# Halos  & \# Halos  \\
& $M_h [10^{14} \hMsun]$&  $M_h [10^{14} \hMsun]$& $z=0$ & $z=0.3$ & $z=0.5$\\\hline
1& 1.00 & 1.12 & 616 & 463 & 352\\
2& 2.00 & 25.0 & 1403 & 759 & 409\\\hline
\end{tabular}
\caption{Mass  bins ranges  for the different redshift
  analysis.
 The last three columns show the number of halos in that bin at a given redshift.}
\label{table:lista_bin}
\end{center}
\end{table*}

\section{Separation Measurement Algorithm}
\label{sec:algorithm}

\begin{figure*}
\begin{center}
\includegraphics[width=0.40\textwidth]{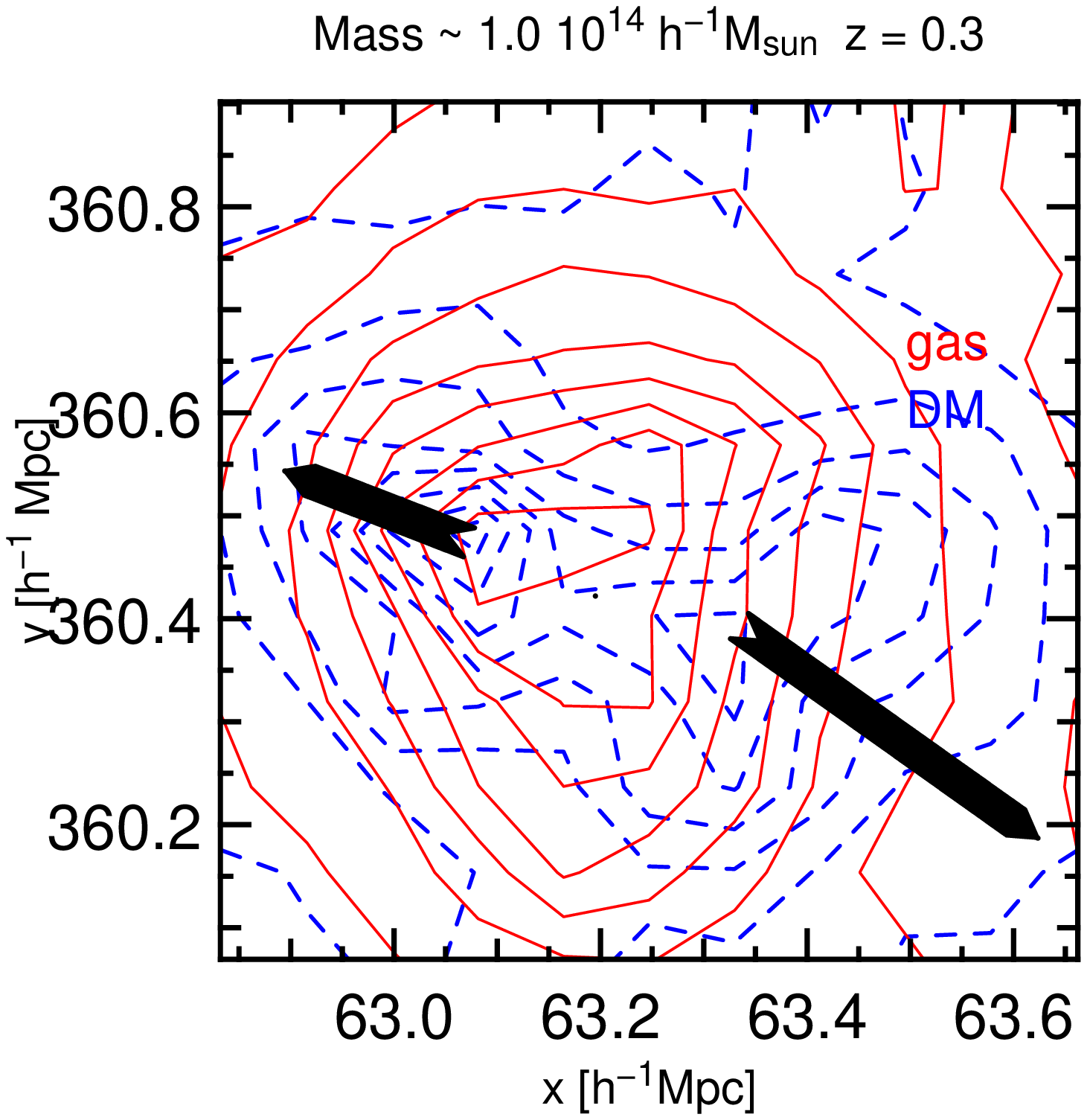}\hspace{0.5cm}
\includegraphics[width=0.38\textwidth]{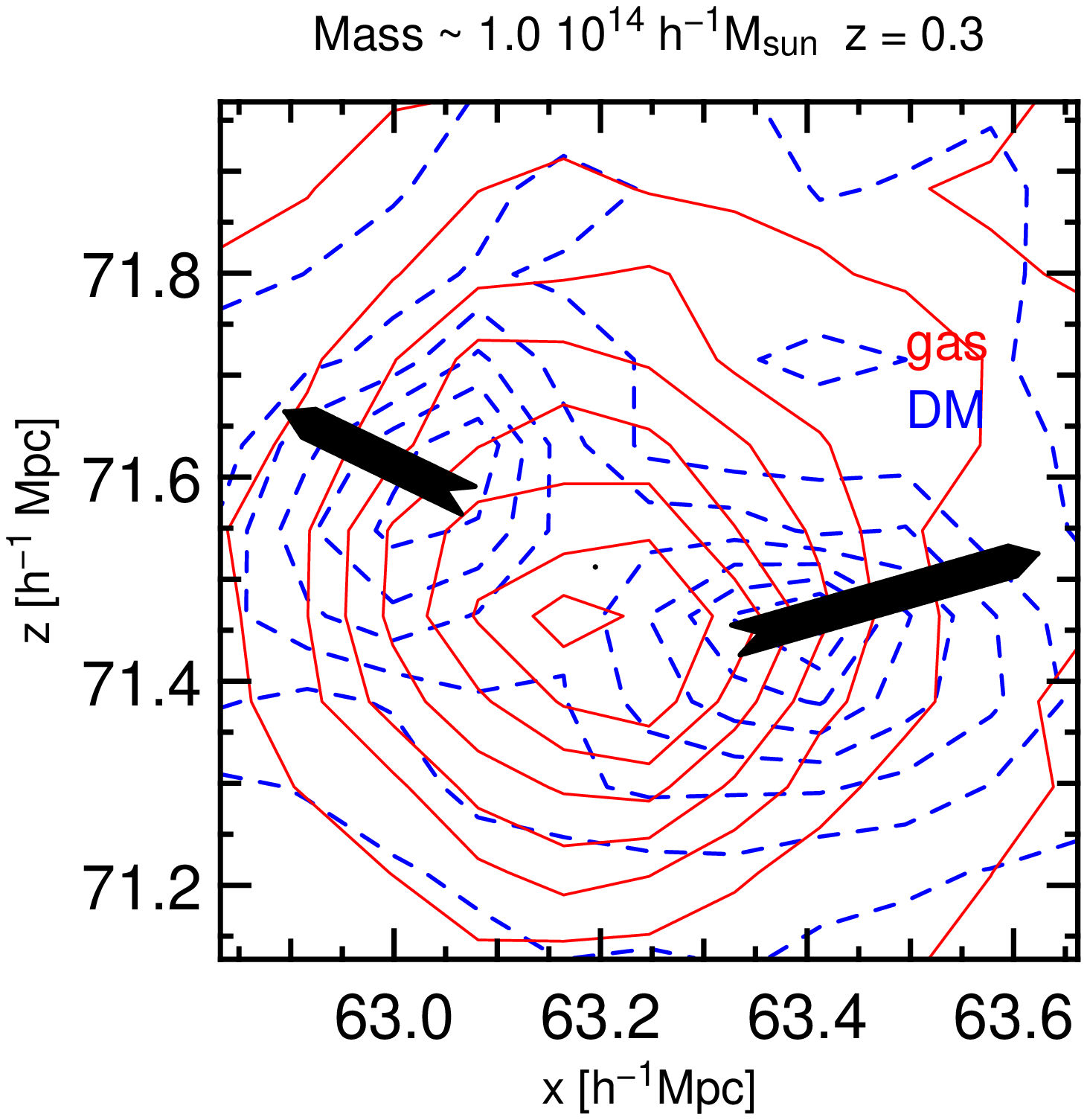}
\end{center}
\caption{Linearly spaced surface density contours for the DM (dashed
  blue) and gas  (continuous red) components of one of the halos  with
  a high displacement seen in two different   projections. The arrows are
  proportional to the velocities respect to the   center of the gas
  core.  The cluster of mass $1.0 \times 10^{14}$ \hMsun is located at redshift $z=0.3$, the coordinates
  are comoving and the relative  speed between the two dark matter
  cores is   560 km/s. In this case, the  physical displacement between
  the  centers   of gas and DM is of the order of  the observed cluster
  \bullb.}     
\label{fig:bullet}
\end{figure*}

The aim of our analysis is to obtain the distribution of displacements
between the dominant density peaks of the DM and gas components of
each selected cluster in the simulation.  We describe first the
algorithm used on a set of hierarchical friend-of-friends (FOF)
catalogs. Then, we describe briefly a different and more complex
algorithm of structure detection,  which, nevertheless,  gives  the
same statistical results than the hierarchical FOF.

\subsection{FOF Based Algorithm}

We have analysed all simulations with the hierarchical
friend-of-friends (FOF) algorithm \citep{1999ApJ...516..530K} used for
the DM and gas particles, separately. The basic FOF groups where
identified using a linking length of $b=0.17$ times the mean inter
particle separation both for the gas and the DM distribution. The
substructures inside the FOF groups are defined as FOF objects
constructed with shorter linking lengths $b_n = b/2^n$ (n=1,2,3).  We
name these catalogs FOF$^{DM}_n$ and FOF$^{gas}_n$ with (n=0,1,2,3)
for the DM and gas components respectively.  For more details on the
simulation, the FOF identification and general cluster properties we
refer the reader to \citet{2007ApJ...664..117G}.
 
For our analysis we have selected all halos with masses larger than
$10^{14}$\hMsun, at four different redshifts, z$=0$, $0.3$ and
$0.5$. The number of selected halos for every redshift, and the
highest halo mass are quoted in Table \ref{table:lista}.

Our objective is to measure the physical separation between the
dominant gas clump in the cluster, with respect to its predominant DM
structure. We find the separation based on the FOF catalogs described
above. The main advantage of the FOF algorithm is that it can be
easily applied to the subsets of DM particles and gas particles and
identifies gas clumps even if they don't have a pronounced density
peak. Moreover, the set of linking lengths in the hierarchical FOF
allows to find objects at different overdensities.  In order to test
the robustness of the results  from the algorithm based on FOF catalogs
 we have compared them with those  obtained using 
  a spherical overdensity algorithm, the Amiga Halo Finder, 
described in the next section.

We define the clusters as the objects detected in the DM FOF$_{0}$
catalog. FOF$_0$ provides objects with about virial overdensity,
whereas FOF$_i$ provides substructures with about $8^i$ higher
densities. In the MN-1024 simulation a typical object of $10^{14}
\hMsun$ at the low mass end of our sample consists of $\sim 10000$ DM
particles and about the same number of gas particles. In the series of
low resolution simulations we consider only clusters with masses
larger than $3.2 \times 10^{14} \hMsun$ which \nobreakspace consists of $\sim 4000$
DM particles and about the same number of gas particles. 

In all cases we find that the clusters exhibit a massive dominant gas
clump and various DM clumps.  In both DM and gas catalogs we define
the dominant clump position in the following way. We identify first
the center and radius of the cluster as given by the DM FOF$_0$
catalog. Next we find the most massive group in the FOF$_1$ catalog
with its center inside the radius defined previously. The clump
defines a new center and radius. We iterate this procedure until we
cannot find a new group in the following FOF$_i$ catalog. The outputs
of the algorithm are the mass and radius of the cluster defined from
the DM FOF$_0$ catalog and the centers of mass of the dominant DM and
gas clumps as well as their separation.

It is well known that the FOF algorithm at the level FOF$_{0}$ will
connect two clusters that are starting to merge as one FOF
object. Since such cases could lead to spurious large separations
between the DM and gas centers they have been excluded from the
analysis. These cases represent always less than $3\%$ of the total
number of clusters.

\subsection{AHF Based Algorithm}

As an example of a spherical overdensity halo finder we have used the
Amiga Halo Finder (\texttt{AHF}) \footnote{It is a MPI+OpenMP hybrid
halo finder to be downloaded freely from {\tt
http://www.popia.ft.uam.es/AMIGA}} which identifies both halos and
subhaloes. \texttt{AHF} locates local overdensities in a adaptively
smoothed grid as prospective halo centers. \texttt{AHF} is described
in detail in \citet{Knollmann2009}. 

The AHF algorithm proceeds by creating a hierarchy of grids. High
refined grids are spanned in high density regions until a minimum
number of particles per grid-cell is reached. The algorithm includes a
criterion to span a new mesh refinement, which is usually expressed as
a density threshold. To find the positions of the gas and DM clumps we
run AHF over the gas and dark matter distribution separately. Thus the
positions can be associated to the regions where the corresponding
mesh refinement reached its highest level. 

In the studied range of separations the two algorithms give the same
answers within the simulation uncertainties. We are confident that our
results are robust for the range of observed Bullet Cluster
displacements.  We proceed the rest of the analysis in this paper
using the hierarchical FOF algorithm, which is less computationally
expensive and much faster than the AHF one.

\section{Cumulative Probability Distributions}
\label{sec:proba}

We found simulated clusters that show larger separations than the
observed Bullet Cluster or the cluster \bullb. We checked by visual
inspection all the cases where the separation between DM and gas peaks
is larger than $200$ kpc. All of them correspond to perturbed
clusters. In Figure \ref{fig:bullet} we show one example. The cluster
was selected because of its striking similarity to the observed
configuration in \bullb.  

We have also checked  the dynamical state of two different samples
  of the clusters at redshift zero.
 The first sample included those with large displacements ($>200$ kpc)
and the second sample  comprises all clusters with low displacement
($<10$ kpc) between gas and dark matter.  We found that, indeed all
clusters with  large displacements come from plunging substructures. 
Moreover,  those clusters which show the most extreme displacements
also present a rich variety of merging configurations, with triple DM
cores in some cases.  On the other hand, in all the cases of  clusters with
  low displaments, the dominant gas   blob has a unique dark matter core
  associated to it, with practically zero   relative velocity between them.  A
  detailed analysis and classification of   all configurations giving rise to
  large displacements in the simulated   cluster will be  presented in a
  forthcoming paper.

From the simulations,  we know the three dimensional positions of the
dominant DM and gas clumps. In order to compare our results with
observations,  we use 2D projections of the 3D displacements into 10
different randomly selected directions. We quantify the results using
the cumulative distribution of the projected 2D displacements.  Using
the 10 different 2D displacement distributions we bin logarithmically
the data in distance and find the mean value and dispersion for the
corresponding cumulative distribution. 

In the following subsections we study the scaling with mass and
redshift of the calculated 2-D displacement cumulative distributions.  

\subsection{Scaling with  Cluster Mass }
\label{subsec:m-scale}

\begin{figure*}
\begin{center}
\includegraphics[width=0.30\textwidth]{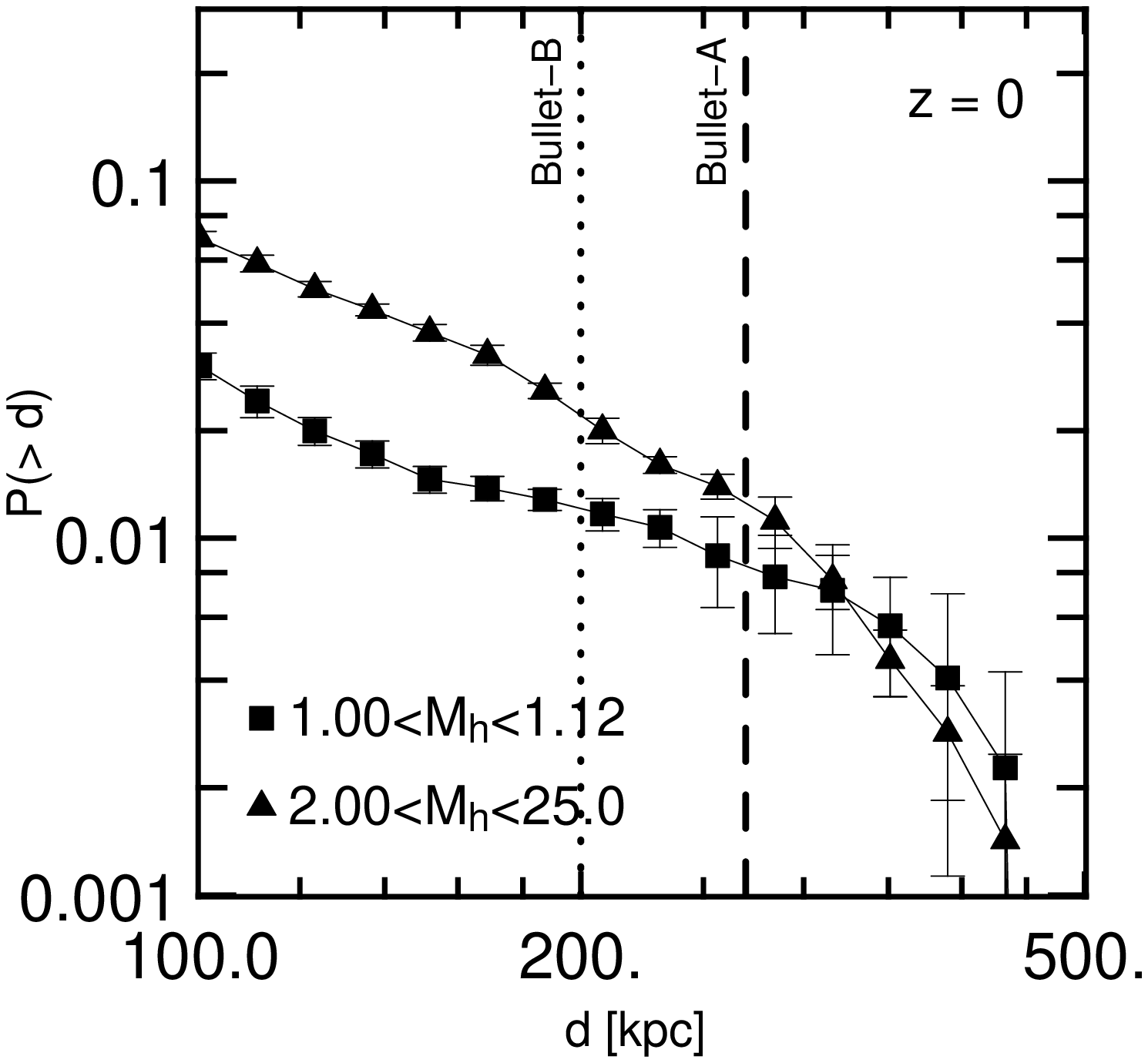}
\includegraphics[width=0.30\textwidth]{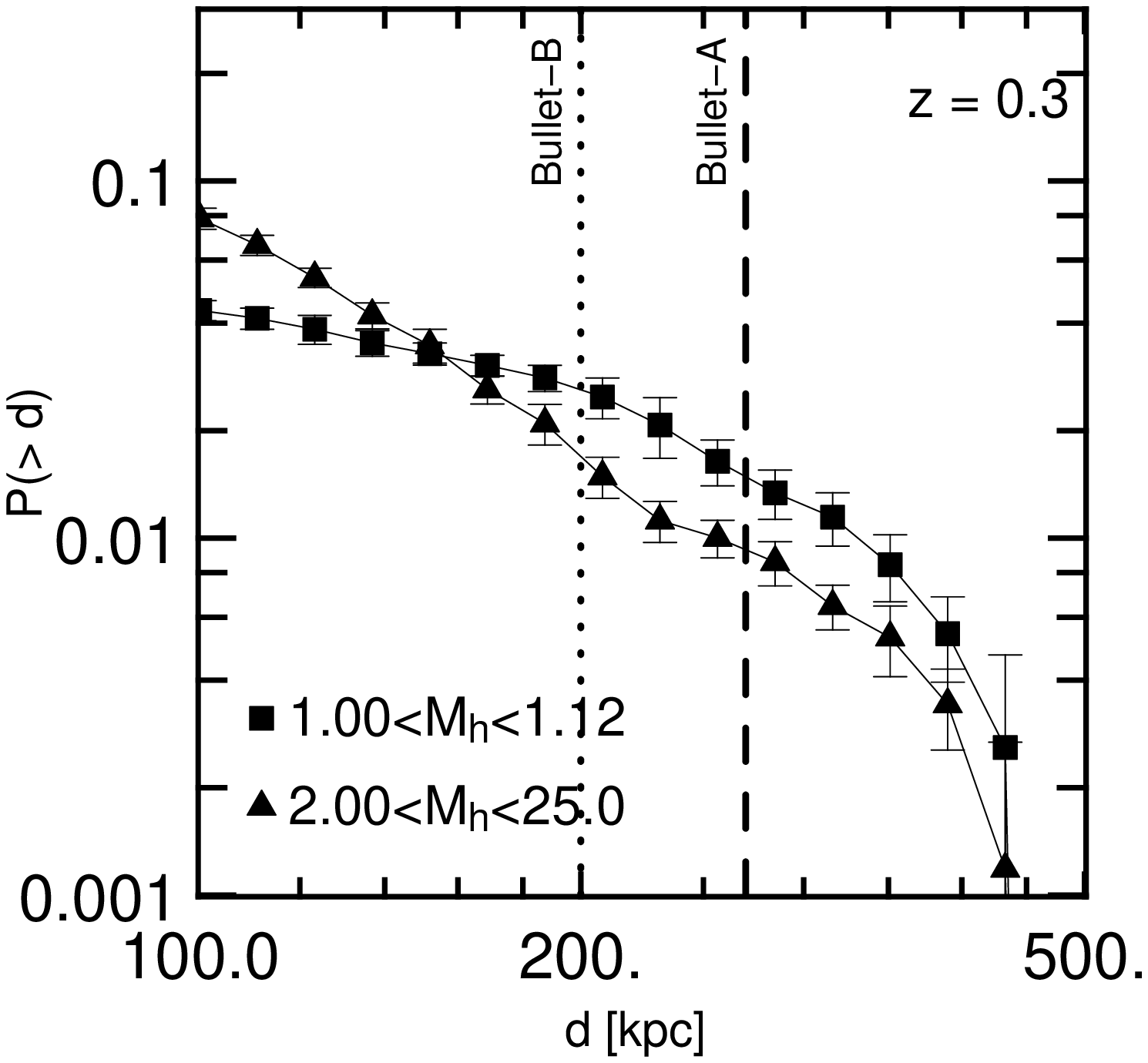}
\includegraphics[width=0.30\textwidth]{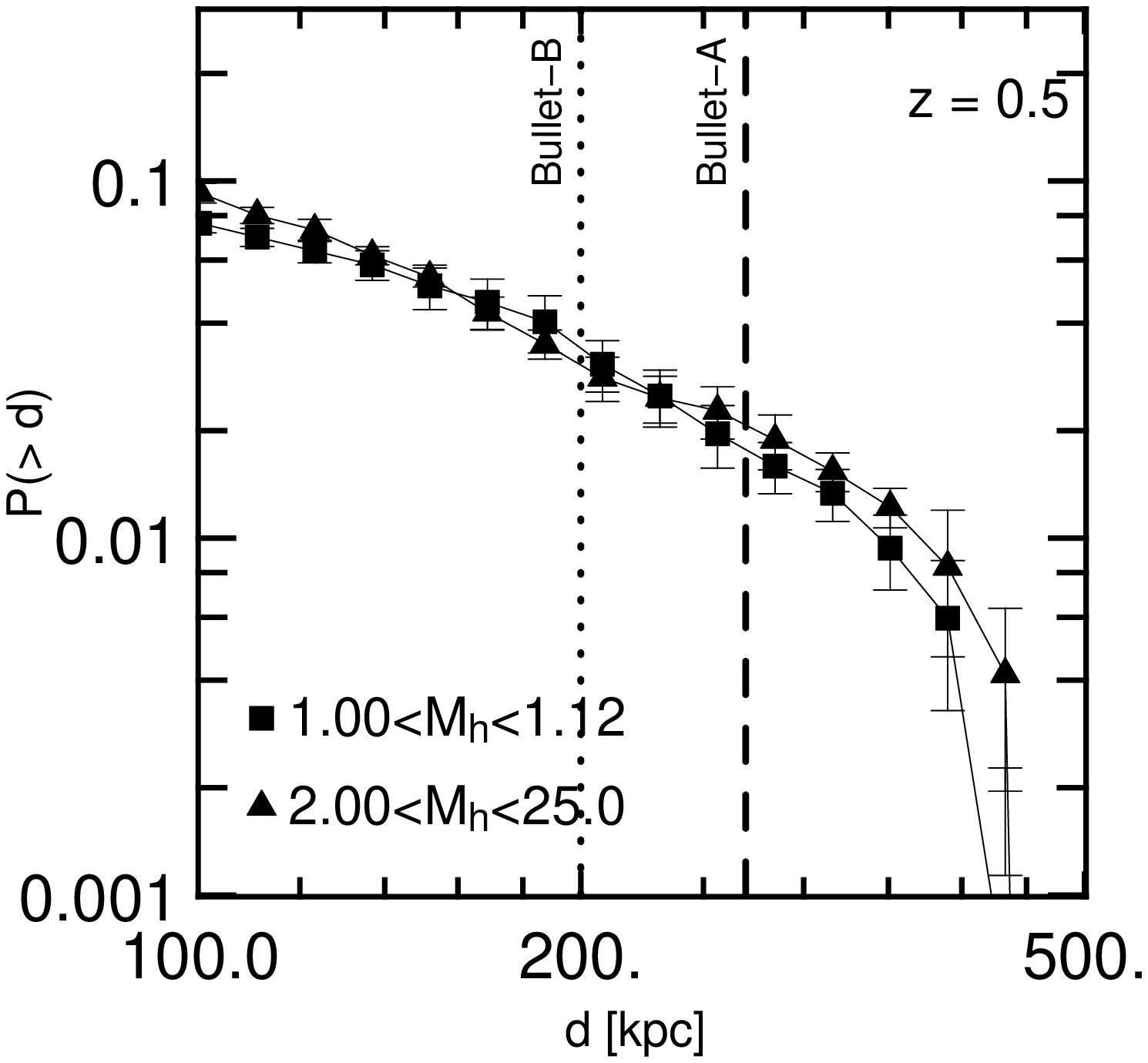}
\end{center}
\caption{Cumulative distribution of 2D displacements between dark
  matter and gas peaks at redshifts $z=0$, $0.3$ and $0.5$ as indicated
  in the corner of each plot. Triangles represent cluster
  more massive than $2.00 \times 10^{14}\hMsun$, squares are clusters with
  masses between $1.00 \times 10^{14}\hMsun$ and $1.12 \times 10^{14}\hMsun$.  
  $M_h$ is given in units of $1\times 10^{14}$ \hMsun. The upper row
  shows the 2D displacements in   physical   units, the lower row
  normalized   by the virial radius of  the cluster. The error bars
  associated to   each point are  calculated   from the variances of
  10 different 2D   cumulative distributions,  computed for different
  projections of the   3D separations measured  in the simulation. The
  vertical lines in the   upper row indicate the  displacements for
  the observed Bullet Cluster   (Bullet-A) and the  cluster \bullb\
  (Bullet-B). The occurrence of a   similar or larger   displacement
  is then to be expected in between 1-2  clusters out of  100. } 
\label{fig:mass-scale}
\end{figure*}

In Figure \ref{fig:mass-scale} we present the cumulative distribution
of projected 2D displacements in the MN-1024 simulation. We bin the
selected clusters in mass, and calculate the distribution for the 2D
physical displacements. The two different mass bins are described in
Table (\ref{table:lista_bin}). The mass bins are selected in such
a way as to have a similar number of clusters in each bin
at $z=0.5$. 
We find
that,  at $z=0$, the fraction of clusters with small  physical
separations  ($\leq 200 $kpc) increases with cluster mass. 
 In constrast, at redshifts
$z=0.3$ and $z=0.5$ we cannot assert that this trend with halo mass
exists at all. In any case,   in the range of the Bullets
displacements, the   distributions  of  different mass bins  are less
than $1\%$ apart for all the redshifts considered here.
As can be seen in the figure, the fraction of simulated clusters with a displacement 
 equal or larger than the observed ones  is always between $1$\% and
 $2$\%. The lowest value corresponds  to the less massive clusters at $z=0$
and the highest value is associated with clusters at high redshifts $z\sim
0.5$.

The statistical properties of the
  displacements at redshift $z=0.3$, for masses larger to $1.0\times
  10^{15}$\hMsun, is nevertheless very poor.  We find a small sample of $8$ clusters. An
  insufficient number to make a solid statistical statement. In spite of that, we found 1 cluster,  from the  sample of $8$ clusters, with
  a    3D displacement larger than the  one measured in the Bullet Cluster.

\subsection{Scaling with Redshift}
\label{subsec:z-scale}

\begin{figure}
\begin{center}
\includegraphics[width=0.45\textwidth]{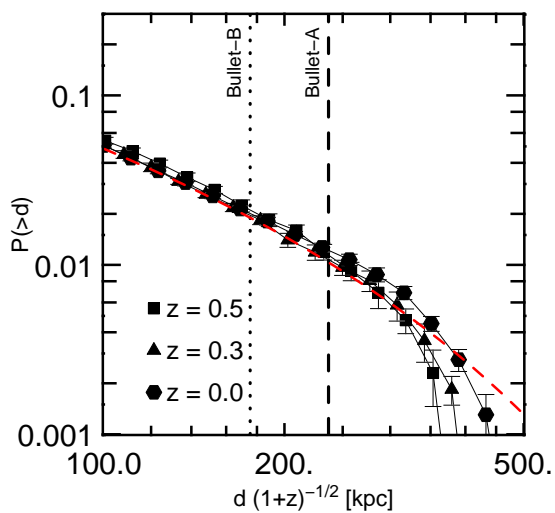}
\end{center}
\caption{Cumulative distribution of 2D physical displacements between dark
  matter and gas peaks  as a function of redshift  scaled by $\sqrt{1+z}$: $z=0$ (hexagons), $0.3$
  (triangles) and $0.5$ (squares). All  halos more
  massive than $10^{14} \hMsun$ were taken into account to make this plot.
 The vertical lines indicate the measured  
  displacements for the observed Bullet Cluster (Bullet-A) and the
  cluster \bullb\ (Bullet-B). 
The red dashed line corresponds to the fitting  formula given  in Eq.(\ref{eq:fit}).} 
\label{fig:z-scale}
\end{figure}

In Figure \ref{fig:z-scale} we show the redshift evolution of the cumulative distribution
of 2D physical  displacements.  All  clusters more massive than $
10^{14}$\hMsun were taken into account  for the  construction of these
distributions. As can be seen in this Figure,  when we consider the
rescaled  variable  $d/\sqrt{1+z}$, where $z$ corresponds to the
redshift, the  cumulative distributions turn out to be the same for
all the redshifts analyzed.

The shape of the cumulative distribution, seems to follow  a power law
of  the displacement until it reaches a critical value,  close to the
Bullets separations $\sim 300$ kpc,  and then falls off exponentially.  A
 Schechter-like function turns out to be  a good fit to the numerical results.  
In terms of the redshift scaled  variable $d_{z}=d/\sqrt{1+z}$:  
\begin{equation}
P_{\mathrm{2D}} =
P_{0}\left(\frac{d_{z}}{d_{\star}}\right)^{\alpha}\exp\left(-\frac{d_{z}}{d_{\star}}\right),
\label{eq:fit}
\end{equation}

with parameters  $P_{0} = 0.04$, $d_{\star} = 200$ kpc and $\alpha=-1.0$, fit
very well the data shown in Figure  \ref{fig:z-scale},  in the range $100<d_{z}/\mathrm{kpc}<500$.

\subsection{A toy model  to explain the  separation dependences}

The redshift scaling of the cumulative probability of displacements
shown  in Fig. \ref{fig:z-scale} calls for an explanation.
To this end we resort to previous numerical studies that have followed
the dynamics of colliding clusters, in a  general context rather than
just focusing on reproducing  the Bullet Cluster.

The are two important studies that we  take as basic references  to this
end.  \cite{2004MNRAS.350.1397T}  followed the properties
of cluster satellites from  hydrodynamical cosmological simulations and 
\cite{2006MNRAS.373..881P}  studied the impact of mergers on relaxed
X-ray clusters, focusing on the dynamical evolution and the emergent
transient structures. Both studies were  based on numerical simulations
which include  gas dynamics, but only the latter included cooling and star
formation as well.  While the work of    \cite{2004MNRAS.350.1397T} is
done in an explicit cosmological setup,   in  the numerical simulations
of  \cite{2006MNRAS.373..881P}  the cosmological context is taken into
account in an implicit manner through an  adequate selection of the initial
conditions. It is relevant to highlight that \cite{2006MNRAS.373..881P}
find that the results from  their  simulations  compare well with  the simple
model predictions in \cite{2004MNRAS.350.1397T}, despite the large
differences in the modelling of the baryonic physics. 

More specifically, 
  \cite{2006MNRAS.373..881P} find that the disrpution of the gas substructure is
quantitative similar to the results obtained by \cite{2004MNRAS.350.1397T}
even though  the effect of cooling allows for the formation of denser
  resilient gas cores. We should also expect that qualitatively, the results
  presented in this paper would not change much if cooling was included in our
  simulation. However, a quantitative statement on how  the displacements we
  measured  are affected by  cooling and star formation processes, would
  require to run  new, high-resolution re-simulations of our clusters. This
  analysis is well beyond the scope of the present work and will be devoted
  in   further publications on the same topic.

In most of the cases studied by \cite{2006MNRAS.373..881P} the central
 ICM of the merger remnant oscillates with respect to the dark matter
 distribution. The oscillation of the main ICM remnant can be seen
 exclusively between the first core interaction and the relaxation of
 the cluster, a process which lasts between $1$Gyr to $3$Gyr depending
 on the mass ratio and the  impact parameter  of the collision. During this
 time interval, the core oscillation presents only one large amplitude
 separation, with the subsequent  separation being around half of the
 first one. A third oscillation shows always an even shorter separation
 on the order of $50$ kpc. One could then simplify the dynamics of the
 gas-DM separation with the description of damped oscillations,
 characterized by an initial amplitude, an intrinsic frequency and a
 damping timescale.  Based on the above results, we try now some
 simple estimations of the scalings  of the physical magnitudes in the
 model.  

Let us consider a momentum conserving collision between two gas
blobs of mass $M_g$ and $m_g$ ($M_g>m_g$), with initial velocities $V_{g}=0$ and
$v_g>0$, respectively. Assuming a totally inelastic collision, the
velocity of the final blob of mass $M_g + m_g$ can be expressed as
$V_{i} = v_{g} m_{g}/ (M_g + m_g)$. With the mass ratio
definition $x=M_g/m_g$, we can rewrite $V_{i} = v_{g}/(x+1)$.
 
N-body simulations  of halo mergers have shown that  the average
velocity of an accreting system into rich cluster type halos  is of the
same order than the circular velocity, $V_c$,  of the main cluster,
independent of its mass
\citep{1997MNRAS.290..411T,2002ApJ...581..799V,2006MNRAS.373..881P}. We
can then approximate $V_{i} \sim V_{c}/(x+1)$.  Thus,  the  first
initial  amplitude of the DM core-core interaction  $A\propto
V_{c}/(x+1)$. 

The intrinsic frequency of the core-core  oscillation can be estimated to be
proportional to the apocentric\footnote{The apocenter is identified as
  the first relative maximum in orbital distance measured after
  the first relative minimum in the satellite orbital distance.} time
for the dark matter component of the satellite. The apocentric time
encodes the information on the typical dynamical time-scale around the
center of the main cluster. \cite{2004MNRAS.350.1397T} find that the
apocentric times depend only on the mass ratio $x$.  

The last element in this description, the intrinsic damping time-scale
can be physically related to the same process that unbinds  gas and dark
matter from the satellite and force their
merging. \cite{2004MNRAS.350.1397T} find that the decay time to unbind
self-bound gas and dark matter particles from the satellite  depends
mainly  on the mass ratio $x$.  

Summarizing, the separation between gas and DM  depends primarily 
 on the circular velocity $V_{c}$ of the main cluster  and the mass
 fraction $x$ of the merging systems.  Thus, the larger amplitudes are
   in principle associated to clusters with larger circular
   velocities, but depend also  on the distribution of mass ratios
   $x$. This might be the reason why we do not see a clear dependence
   with cluster  mass  at all redshifts.

 Moreover, there is also  a redshift dependent scaling relationship
   between the circular velocity $V_{c}$ in physical magnitudes, and
   the halo mass $M_{h}$. 
 The redshift scaling is such that $d \propto V_{c} \propto M_{h}^{0.35}\sqrt{1+z}$
\citep{2004MNRAS.350.1397T}.

In Section \ref{subsec:z-scale} we considered the displacements of all
 halos. In this case, the cumulative probability distribution is
dominated by the low mass halos simply because of its higher abundance,
meaning that we are probing a similar mass range $\sim 10^{14}$\hMsun at
different redshifts. From the numerical data we find that the distribution of
$d/\sqrt{1+z}$ is redshift independent. This redshift scaling is consistent with the
one  derived from our toy model,  $d \propto M_{h}^{0.35}\sqrt{1+z}$.

\subsection{How many Bullets should we expect?}

One can now ask the following question: which is the
fraction of clusters that will show a separation larger than a
given distance? If the volume limited sample is sensitive to halos
down to $10^{14}$ \hMsun, the answer to this question is given by
Figure \ref{fig:z-scale}:  a fraction between  $1\%$ and $2\%$ of the clusters
presents  a separation equal or larger than the Bullet, depending  mostly
on the redshift, and weakly on the cluster mass,   as seen in Figure \ref{fig:mass-scale}. If we   extrapolate  these results to  clusters with masses larger than in   our simulations,  we would expect  that the fraction of bullet-like   displacements will not be very much different  (within a factor of $\sim  2$)   of the  1-2 \% shown here.

\section{Dependence on Resolution and Power Spectrum Normalization}
\label{sec:power}

We will quantify now the dependence of our results on the numerical
resolution of the simulations.
 To this end we compare the cumulative
distribution of 2D displacements of DM and gas centers at redshift
$z=0$ of the simulations MN-1024 and MN-512.  The only difference
between the two simulations is the number of particles and the
gravitational smoothing.  The same seed has been used for 
  the initial conditions. 

\begin{figure*}
\begin{center}
\includegraphics[width=0.4\textwidth]{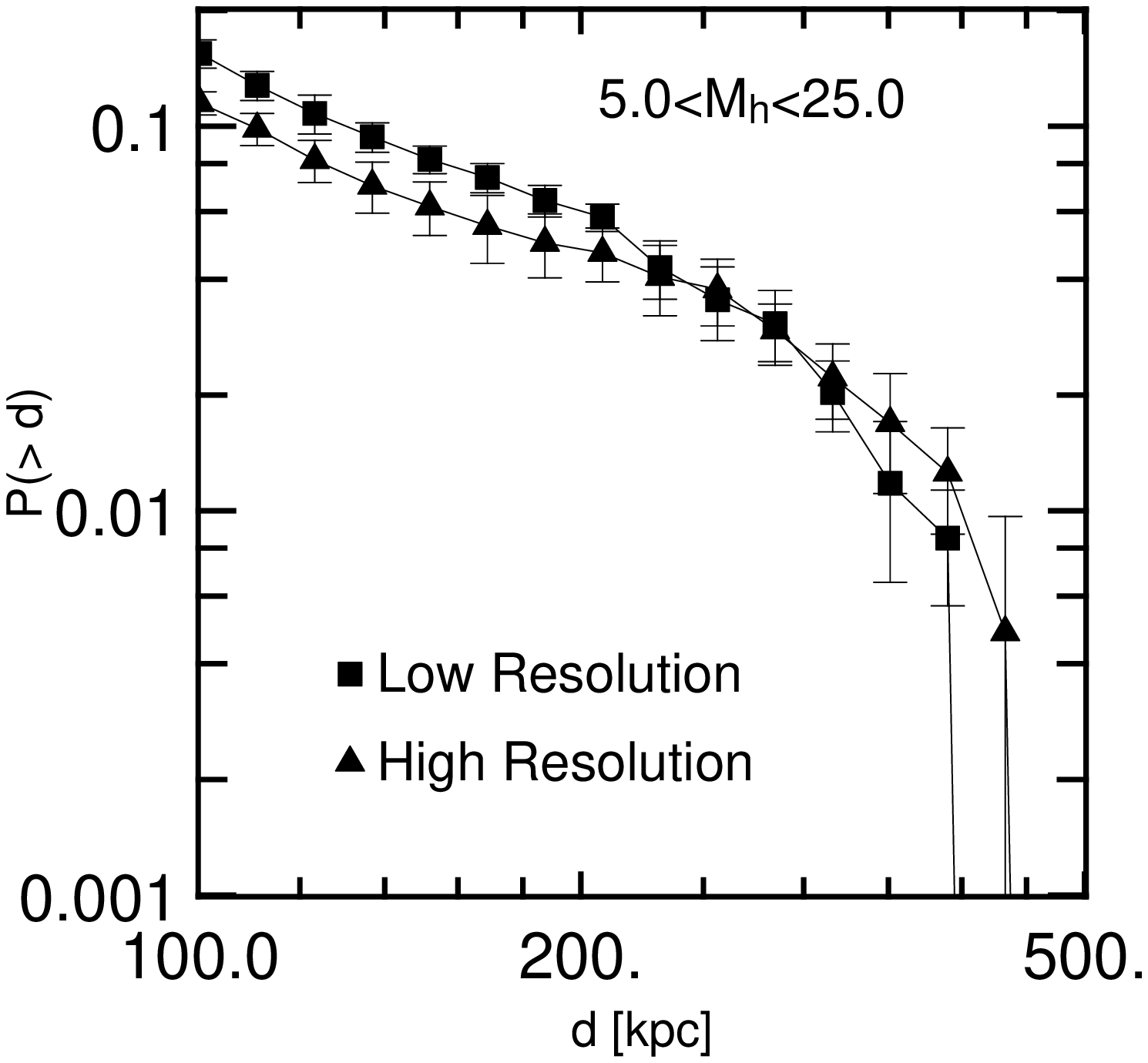}
\includegraphics[width=0.4\textwidth]{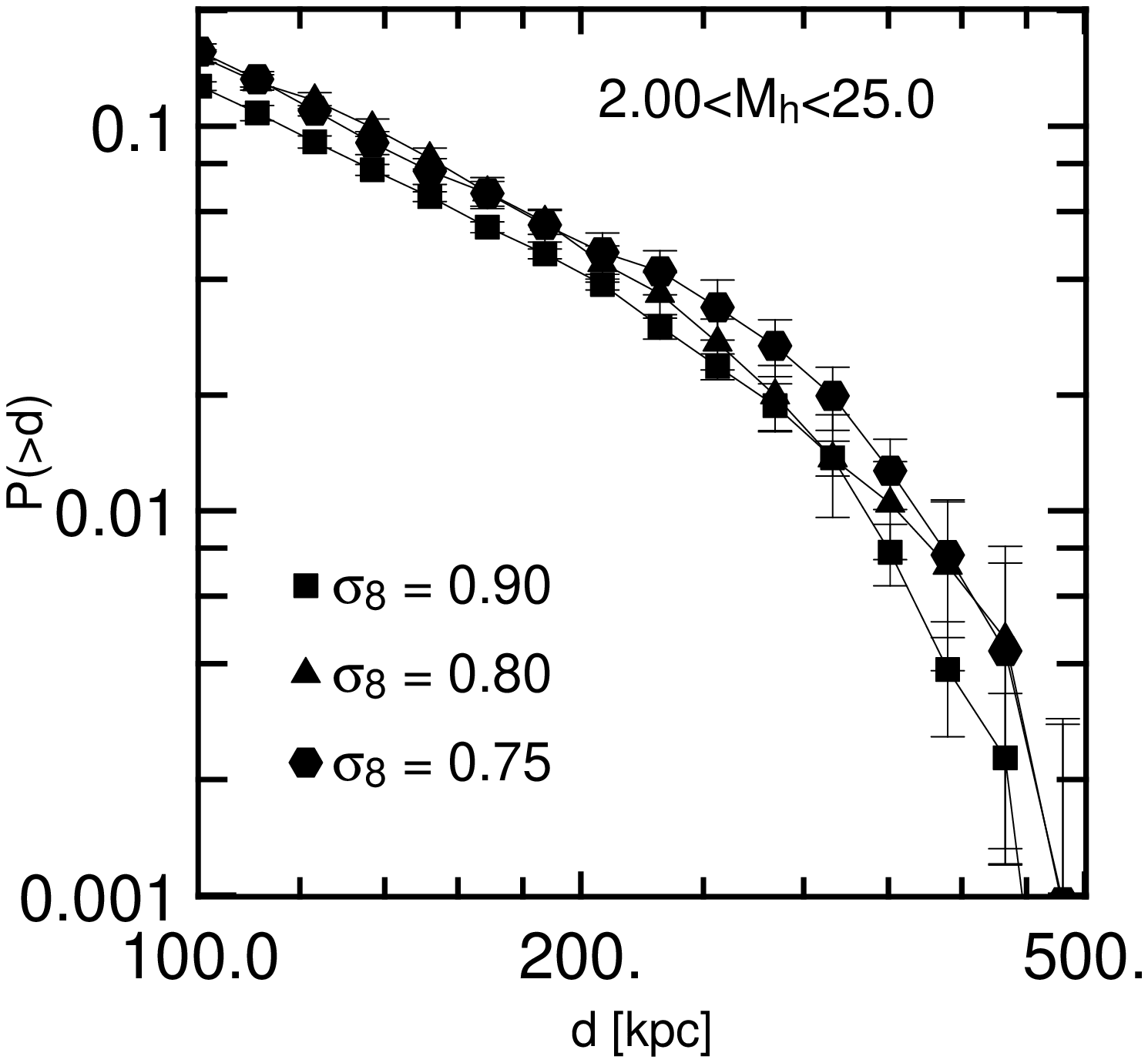}
\end{center}
\caption{Left panel: Cumulative distributions of 2D displacements for
  simulations with different mass resolutions. Righ panel:  Cumulative
  distributions of 2D displacements for simulations with different
  power spectrum normalizations $\sigma_8$. $M_h$ is given in $10^{14}
  \hMsun$.}
\label{fig:res_sigma}
\end{figure*}

For this comparison we use the highest mass bin in which  we have, 
in the low resolution simulation,  about 4000 DM particles per cluster.
This is  enough to determine the density peaks with high precision. In
the left panel of Figure \ref{fig:res_sigma} we show that there is 
a good agreement  between  the cumulative probability distribution
 of the DM-gas displacements in the two simulations with different
 resolution.  

Therefore, we are
convinced that our results are  not likely to be affected by numerical
resolution as long as we have 10.000 or more gas and DM particles available in the clusters under consideration.

Now,  we want to use the low resolution simulations to study the
dependence of our predictions on the underlying cosmological model. Here we focus
again on clusters with more than 4000 DM and gas particles. We compare
three models with different normalisation and matter content which are
roughly compatible with the WMAP1, 3, and 5 predictions (see Table
\ref{table:sims}). \citet{2007ApJ...666L..61Y} have shown that different
$\sigma_8$ give rise to strong variations  in X-ray cluster abundance.

We repeat the analysis presented above using the low resolution
simulations MN-512-LS, MN-512 and MN-512-MS. In the right panel of
Fig. \ref{fig:res_sigma} we show the 2D probability separation
distributions for the three lower resolution  simulations at redshift
$z=0$. We cannot argue for a significant difference in the probability
distributions derived from simulations with  different $\sigma_8$ .
 Although  the total number of cluster   is different in each
 simulation,   the fraction of clusters showing a give
displacement of DM-gas peaks   remains almost unchanged.

\section{Are Bullet Clusters a Challenge to $\Lambda$CDM?}
\label{sec:challenge}

The recent work by \cite{lee} claims again the difficulty represented
by the Bullet Cluster to the $\Lambda$CDM cosmology. Their analysis is
very similar in spirit to the work of \cite{Hayashi06}. \cite{lee}
analyze a large volume dark matter only simulation looking for high
mass clusters that were not found in the relatively small volume
spanned by the Millennium Simulation. The relative velocity of the
structure colliding with the main cluster is taken from the study of
\cite{2008MNRAS.389..967M}.

The study undertaken by \cite{2008MNRAS.389..967M} explores the
parameter space in a set of idealized cluster collisions.  The best
mass ratio, impact parameter, and relative velocities of the collision
are constrained by assuming that a set of four properties is
reasonably reproduced: i) the displacements between the X-ray peaks
and the mass distribution, ii) the morphology of the bow shock, iii)
brightness of the shock and iv) the morphology of the main gas
peak. Most of the simulations were performed with a non-radiative
treatment of the gas physics.  Two different runs out of the total 13
included a crude implementation for cooling. Leaving aside the well
known overcooling problem, which affects the main gas peak, the
brightness of the shock is strongly modified. Thus the simulated
properties of the bow shock are still not well understood and are not
yet robustly described by these simulations. 

One of the most robust simulated properties is related to the main gas
peak.  It is well known that a model of cooling without inclusion of
energetic feedback does not give a proper description of the gas in
the inner part of a cluster \citep{2009arXiv0906.4370B}. Nevertheless,
already the non-radiative description of the gas is a reasonable
approximation for the dynamics of the main gas peak
\citep{2006MNRAS.373..881P}. 
A complete description of the physics of shocks, should include
cooling, star formation and feedback.  The fact that the bow shock
properties are not robustly described neither in adiabatic nor in
radiative runs, weakens the case for its selection as one of the
criteria to constrain the success of a certain cosmological model in
reproducing the observations. We doubt that the disagreement of the
predicted and observed properties of colliding clusters depends only
on the parameters of the collisions which are determined by the
underlying $\Lambda$CDM model.

In other words, simulations of structure formation in the $\Lambda$CDM
paradigm still do not predict robustly the present very detailed X-ray
observations of colliding clusters. Therefore, at present it it is
difficult (or impossible) to decide whether or not the complete set of
observed properties of the cluster \bulla\ show a significant anomaly.

In this paper we have taken a much more modest objective, of selecting
a single robust property in $\Lambda$CDM simulations of clusters,
namely the distance between the main dark matter and gas peaks, to
work out the significance of the observed anomalies in clusters like
\bulla\ and \bullb\ . We find that large DM-gas displacements are
fairly common in a $\Lambda$CDM cosmology. Furthermore our theoretical
estimates of the fraction of clusters with large displacements are
roughly consistent with the recent observational study of \cite{shan}.
Therefore, we consider that Bullet Clusters should not be considered
as a challenge to $\Lambda$CDM.

\section{Summary and Conclusions}
\label{sec:conclusions}

Inspired by the displacement between gaseous and dark component
observed in a number of clusters of galaxies, we calculate the
distribution of these displacements in a large cosmological SPH simulation
including dark matter and gas.  

Theoretically, the mechanism giving rise to these displacements
involves a high velocity substructure passing through the center of the
parent cluster halo. Assessing the feasibility to accommodate that kind
of events in the $\Lambda$CDM context, could be used as a test for the
validity of the model. Unfortunately, deriving  the relative velocity from 
observations is a non-trivial task, because of statistical and
systematical uncertainties, making difficult the comparison to
theoretical models.  

The separation between the DM and gas components is a much better
defined quantity, mostly affected by statistical errors in the
measurement process, but not so much by systematic  or model-dependent
uncertainties.  

In this paper  we derive,  for the first time,  the expected physical
2D separation distribution for clusters with a DM mass larger than
$ 10^{14}$\hMsun. We find that around $1\%$ to $2\%$ of these
clusters show DM-gas separations equal or larger than the observed in
the Bullet Cluster.  Thus,  the existence of Bullet Clusters should not
be considered as a challenge to the $\Lambda$CDM model as it has been
recently claimed.

Even though  the fractions of expected bullets in  our $\Lambda$CDM
  simulation  basically  coincide with the  previous    work of
  \cite{Hayashi06}, the comparison  between the two  results   should be taken
  with a bit of caution.  As we mentioned before, the definition of  a  Bullet
  cluster  in \cite{Hayashi06}   is different than in our case.   Moreover,
  the dynamical origin of  the large displacements   in our  clusters seems to
  be more  varied than   the scenario of a single dominant substructure
  passing through the cluster.  Nevertheless,  the striking coincidence of the
  predicted bullet  fractions  between the Millenium N-body simulation and the
  MareNostrum SPH simulation suggests that indeed it is possible to produce a
  bullet-like   configuration with substructure velocities as low as $\sim
  2400$ km/s. 

The approach we use and the results we found are above all a necessary
complement to the work of \citet{Hayashi06} and \citet{lee}. We give
information about the configuration space rather than the velocity
space, which can be compared in a more straightforward way to
observations as it has been done in the recent work of \cite{shan} on
a small sample of 38 clusters, who find an acceptable agreement with
our theoretical predictions. The full distribution of these
displacements can be considered as new prediction of the $\Lambda$CDM
model, which could eventually be compared against
observations. Upcoming large optical and X-rays surveys can make this
feasible in the next years.

\section*{Acknowledgements}

The authors acknowledge the work of Claudio Llinares in early stages
of the project and Alexander Knebe for his help in setting up his AHF
code to be used in our simulations. The {\sc MareNostrum Universe}
simulation has been done at the Barcelona Supercomputer Center and
analyzed at NIC Juelich.  GY acknowledges support of MEC (Spain)
through research grants FPA2009-08958 and AYA2009-13875-C03-02. We
also acknowledge the CONSOLIDER-INGENIO projects MULTIDARK
(CSD2009-00064) and SyEC (CSD2007-0050) for supporting our
collaboration. We would also like to thank the anonymous referee for her/his
valuable comments.

\end{document}